\documentclass[10pt,conference]{IEEEtran}
\IEEEoverridecommandlockouts
\usepackage[left=0.6in,right=0.6in,top=0.75in,bottom=0.75in]{geometry}
\usepackage{cite}
\usepackage{amsmath,amssymb,amsfonts}
\usepackage{algorithmic}
\usepackage{graphicx}
\usepackage{textcomp}
\usepackage{xcolor}
\usepackage{mathtools, nccmath}

\usepackage{hyperref}
\usepackage{caption}
\usepackage{subcaption}
\usepackage{booktabs} 
\usepackage{makecell} 
\usepackage{multirow}
\usepackage[version=4]{mhchem} 
\usepackage{url}
\usepackage{algorithm}
\usepackage[final]{microtype}
\usepackage{soul} 

\newcommand{\squeezeupf}{\vspace{-5mm}}
\newcommand{\squeezeupt}{\vspace{-2mm}}

\def\BibTeX{{\rm B\kern-.05em{\sc i\kern-.025em b}\kern-.08em
    T\kern-.1667em\lower.7ex\hbox{E}\kern-.125emX}}
\begin{document}
\title{True Random Number Generation using Latency Variations of Commercial MRAM Chips}

\newcommand*{\affmark}[1][*]{\textsuperscript{#1}}
\author{\IEEEauthorblockN{Farah Ferdaus\affmark[$\ast$],
B. M. S. Bahar Talukder\affmark[$\dagger$], 
Mehdi Sadi\affmark[$\mathsection$], and 
Md Tauhidur Rahman\affmark[$\ddagger$]}
\IEEEauthorblockA{\affmark[$\ast\dagger\ddagger$]\textit{ECE Department, Florida International University, Miami, FL, USA} \\
\affmark[$\mathsection$]\textit{ECE Department, Auburn University, Auburn, AL, USA} \\
\{\affmark[$\ast$]fferd006, \affmark[$\dagger$]bbaha007,  \affmark[$\ddagger$]mdtrahma\}@fiu.edu, \affmark[$\mathsection$]mehdi.sadi@auburn.edu}
\vspace{-8mm}
}
\maketitle

\begin{abstract}

The emerging magneto-resistive RAM (MRAM) has considerable potential to become a universal memory technology because of its several advantages: unlimited endurance, lower read/write latency, ultralow-power operation, high-density, and CMOS compatibility, etc. This paper will demonstrate an effective technique to generate random numbers from energy-efficient consumer-off-the-shelf (COTS) MRAM chips. In the proposed scheme, the inherent (intrinsic/extrinsic process variation) stochastic switching behavior of magnetic tunnel junctions (MTJs) is exploited by manipulating the write latency of COTS MRAM chips. This is the first system-level experimental implementation of true random number generator (TRNG) using COTS toggle MRAM technology to the best of our knowledge. The experimental results and subsequent NIST SP-800-22 suite test reveal that the proposed latency-based TRNG is acceptably fast ($\sim 22 Mbit/s$ in the worst case) and robust over a wide range of operating conditions.

\end{abstract}


\section{Introduction} \label{sec:intro}
True random number generator (TRNG) plays an important role in cryptographic applications such as random key generation, cryptographic nonces, session keys, one-time-pad, initial seeds of pseudo-random number generator (PRNG), challenges for authentication, hardware metering, etc. \cite{trng_Mathew, trng_kim, trng_dram_kim}. A TRNG translates random physical phenomena (i.e., physical entropy) into digital sequences. Thermal noise of resistors and capacitors \cite{trng_kim,TRNG_rahman_sram}, meta-stability \cite{trng_yang, trng_mStability}, random telegraph noise in dielectrics \cite{rtn_Brederlow,trng_reram}, oscillator jitter \cite{trng_kim,TRNG_rahman}, chaos \cite{trng_kim}, quantum phenomena \cite{trng_quantum}, random spintronic \cite{Spin_ghosh,mram_trng_choi,Spin_chun} and memristive \cite{reram_trng_wei,reram_trng_Balatti} properties, atmospheric-, shot-, radio- and flicker-noise \cite{trng_sara, trng_dram_kim}, etc. are the most common high-quality physical entropy sources that are harvested to generate random numbers. 
Furthermore, the process variation during integrated circuits (ICs) fabrication is also responsible for random noise \cite{proc_var}. In most cryptographic applications, the quality of system's security relies on the quality of random numbers. A poor TRNG can always be a target to an adversary for attacking the whole system. A TRNG can be a discrete or an integral part of the system (e.g., on-chip TRNG). Usually, an on-chip TRNG has several advantages: low-overhead (area and energy), non-deterministic, high-throughput, simple design, robust against a wide range of operating conditions, etc.

Continual scaling down in technology introduces enormous challenges such as substantial process variation, crucial transistor's sensitivity with different operating conditions, significant power consumption, etc. to the existing memory chips. Current mainstream volatile memory chips, i.e., static RAM (SRAM) and dynamic RAM (DRAM), suffer from scalability, density, memory persistency, and leakage issues. On the other hand, existing non-volatile memory (NVM) chips (e.g., Flash) suffer from performance and endurance problems. Due to these limitations, existing memory chips are incompetent in delivering ever-increasing demands of power-efficient, smaller, and high-performance systems \cite{DRAM_limit}. Thankfully, MRAM can turn into a dominant universal memory (cache and main memory) technology due to its promising scopes such as non-volatility, scalability, unlimited endurance, high speed, and fast read access, ultralow-power operation, CMOS compatibility, reliability, high density, thermal robustness, and radiation hardness \cite{MRAM_propty}. Because of these advantages, most of the systems are expected to include MRAM chips. Therefore, MRAM can be an attractive candidate for low-power TRNG.

There have been several high-quality and robust memory-based TRNGs, but most of them suffer from high-overhead and low-throughput \cite{trng_flash, trng_Mathew, trng_kim, mram_trng_yang, trng_reram, trng_reram2}. Therefore, several emerging memory-based TRNGs, capable of providing high density and throughput, have been proposed to overcome existing challenges \cite{mram_trng_choi, reram_trng_Balatti, FRAM, trng_reram, trng_reram2}. Furthermore, MRAM-based TRNGs have gained attention because of their capability of generating significantly high quality and robust random numbers \cite{trng_khan, mram_trng_choi, mram_trng_yang, mram_puf_trng}. However, the existing MRAM-based TRNGs are mostly simulation-based or need modification in standard MRAM structures \cite{mram_puf_trng, mram_trng_Fukushima,mram_trng_yang}. Furthermore, some MRAM-based TRNGs can not be easily integrated into the existing computing system due to the strict requirement on operating conditions (e.g., precise control over current/voltage pulse width/magnitude/waveform). \cite{mram_trng_choi, mram_trng_Carboni, trng_khan}. 

The previous contributions inspirit the need for real memory implementation and build the foundation of proposed MRAM-based TRNGs generated from COTS MRAM chips, which require minimal or no additional hardware, are robust against environmental fluctuations and provide considerably high throughput. In this work, we propose a technique of generating random numbers that meet the aforementioned requirements by exploiting write latency variations of COTS MRAM chips. In summary, the major contributions of this work are as follows.
\begin{itemize}
\item {We reduce the write enable ($\overline{W}$) time from the manufacturer recommended value during the write operation to introduce errors. Errors from some of the cells at the reduced timing parameter are entirely random and can be used as a source of randomness. However, some of the cells exhibit deterministic behavior. Therefore, we further propose an algorithm to select the most suitable memory cells that exhibit proper randomness to generate robust and high-quality random numbers.}
\item {We demonstrate the system throughput and robustness of our proposed TRNG in multiple COTS Everspin toggle MRAM chips (\cite{everspin}) under a wide range of operating conditions.}
\end{itemize}

The rest of the paper is organized as follows. Sect. \ref{sec:mram} briefly overviews the organization and operating principle of MRAM chips. Sect. \ref{sec:method} presents the proposed technique of generating true random numbers, including cell characterization and suitable bit-selection algorithm. Sect. \ref{sec:result} explains the experimental setup and exhibits obtained results to verify the quality and robustness of the proposed TRNG. Finally, Sect. \ref{sec:end} concludes the paper.

\section{MRAM Architecture and Operation} \label{sec:mram}

\begin{figure*}[ht!]
    \centering
    \captionsetup{justification=centering, margin= 0.5cm}
    \begin{subfigure}[t]{0.3\textwidth}
        \centering
        \includegraphics[trim=0cm 10.5cm 20cm 0cm, clip, width=0.9\textwidth]{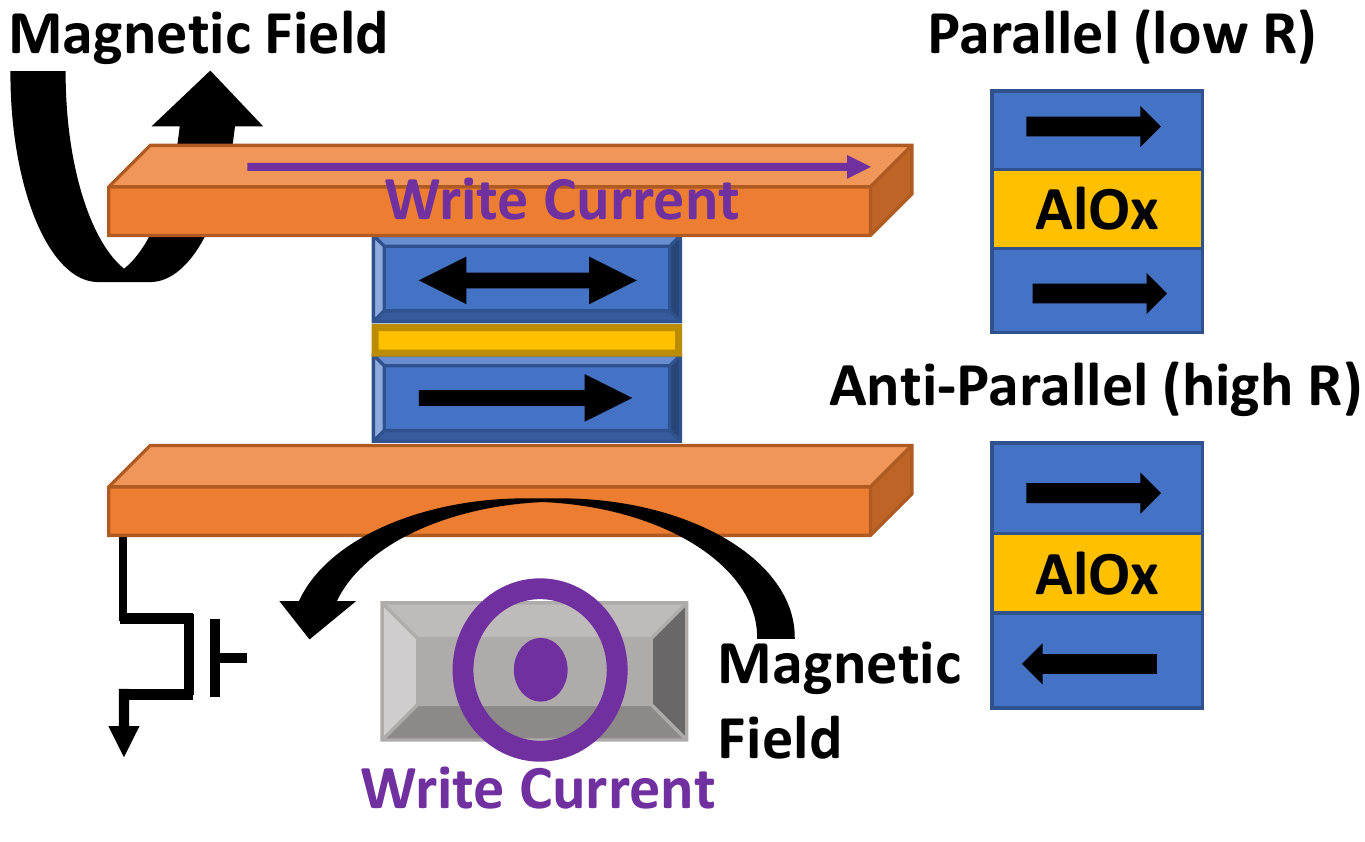}
        \caption{}
        \label{fig:MRAMcell}
    \end{subfigure}%
    \begin{subfigure}[t]{0.32\textwidth}
        \centering
        \includegraphics[trim=0cm 7.5cm 14cm 0cm, clip, width = 0.9\textwidth]{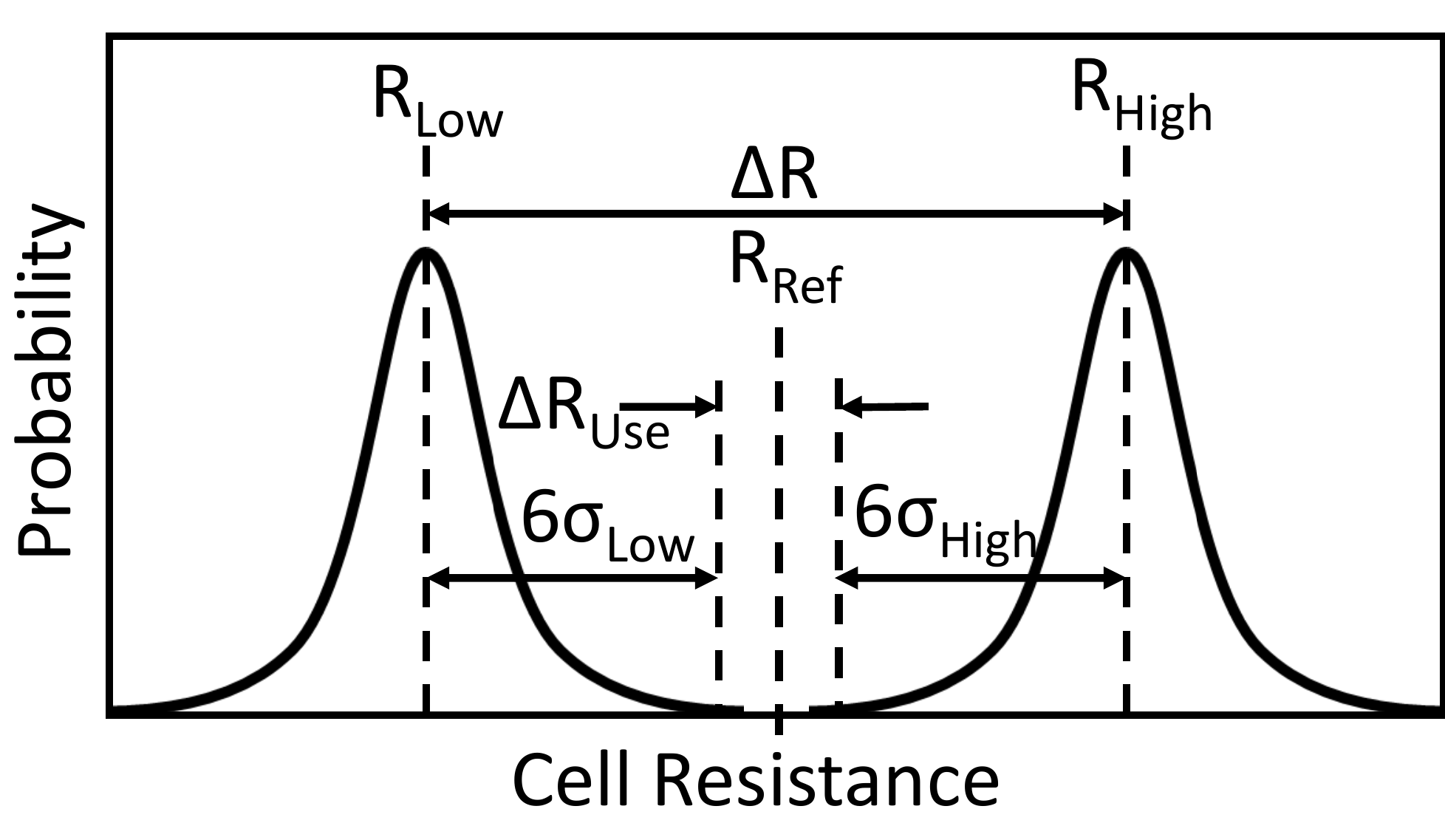}
        \caption{}
        \label{fig:cell_res}
    \end{subfigure}%
    \begin{subfigure}[t]{0.35\textwidth}
        \centering
        \includegraphics[trim=0cm 10cm 10cm 0.5cm, clip, width = 0.9\textwidth]{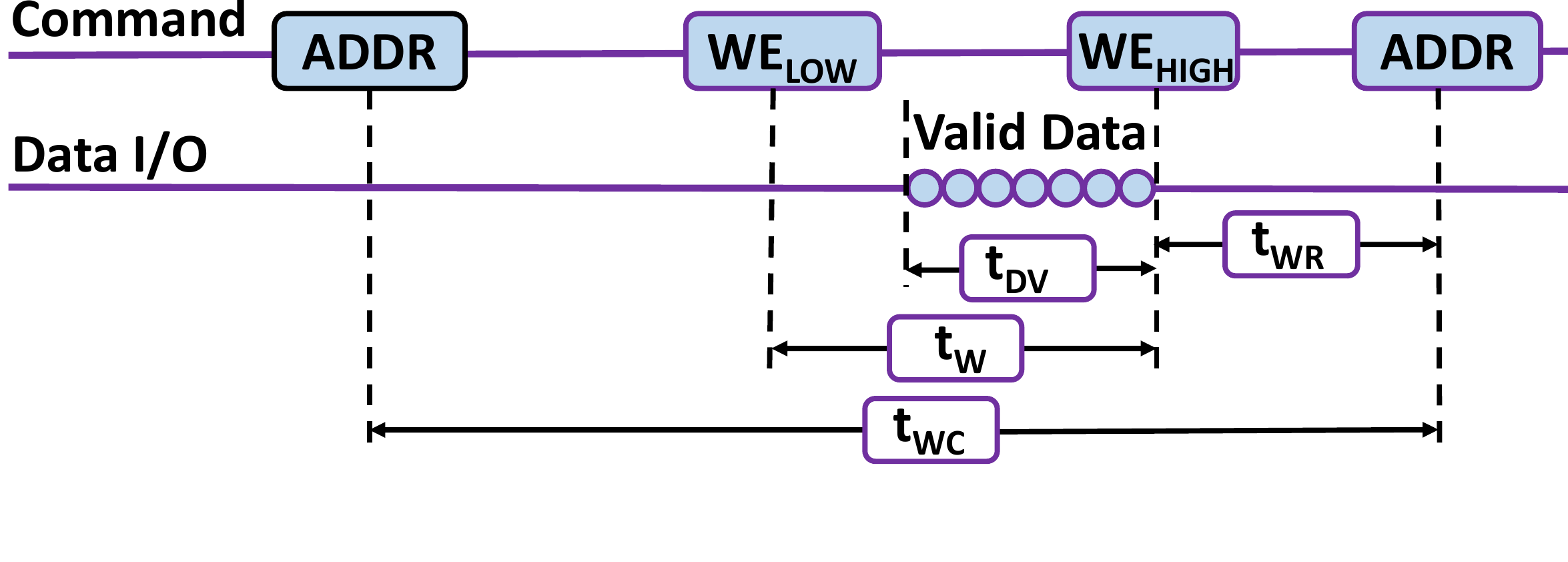}
        \caption{}
        \label{fig:Everspin}
    \end{subfigure}
    \caption{(a) Toggle MRAM cell structure with MTJ. (b) Schematic representation of Gaussian resistance ($R_{Low}$ and $R_{High}$ states) distribution of larger-sized MTJ array \cite{toggle_mram2}, and (c) Write enable ($\overline{W}$) controlled write cycle of MRAM chip.}
    \label{fig:MRAM}
\squeezeupf
\end{figure*}

Magnetic tunnel junction (MTJ) is the core element of toggle MRAM that uses
the Savtchenko switching \cite{toggle_mram, toggle_mram2} property by creating a rotating field with the sequential identical
write current pulses to store both (high and low) data states. The bit cell of 1T-1MTJ MRAM architecture comprises two ferromagnetic layers separated by a thin dielectric tunnel oxide ($\ce{AlOx}$ or $\ce{MgO}$) layer (shown in Fig. \ref{fig:MRAMcell}). One layer’s magnetic orientation is always fixed, known as the reference (or fixed) magnetic layer (RML). Depending on the magnetic field, another layer’s magnetization can freely be oriented, and this layer is known as the free magnetic layer (FML). The FML is composed of $\ce{NiFe}$ synthetic antiferromagnet (SAF). The considerably higher magnetic anisotropy of RML compared to FML ensures stable magnetization direction of FML during memory (read/write) operation. Storing bits in the memory array is determined by the resistance states. When both the FML and RML are aligned in the same direction (current passed from SelectLine (SL) to BitLine (BL)), the MTJ produces low electrical resistance. 
On the other hand, when their magnetic field orientation is opposite, the MTJ exhibits high electrical resistance.

%

Writing bits in the magnetic field-driven toggle MRAM array requires passing a high write current ($I_w$) for changing FML’s magnetic orientation~\cite{toggle_mram2}. The applied $I_w$ to 
the write lines, placed on top and bottom of the MTJ devices (see Fig. \ref{fig:MRAMcell}) creates an auxiliary magnetic field that changes FML direction. On the other hand, the direction of RML is strongly coupled with an anti-ferromagnet \cite{toggle_mram2}. During the write operation, the memory circuit performs a pre-read operation to determine the state of the target bit and execute a toggle pulse (if required) to change the state of the bit if the desired state is not the same as the target state. Consequently, it reduces the overall power consumption and improves power efficiency. However, this increases the total write cycle time (including an additional read operation).

During the read cycle, a small bias voltage (far below the breakdown voltage of the device) is applied across the MRAM cell. Depending on parallel ($R_{Low}$) or anti-parallel ($R_{High}$) orientation, a current sensing circuitry (attached with the MRAM cell) experiences different current and latches the appropriate logic (`0' or `1') comparing with the reference resistance ($R_{Ref}$) shown in Fig. \ref{fig:cell_res}. Fig. \ref{fig:cell_res} illustrates the random resistance variation effect of the read circuitry of a larger-sized MRAM array. Those bits are considered acceptable if their statistical separation is greater than $5\sigma$ from the mean, where $\sigma$ is the standard deviation. The accuracy of the read circuitry depends on determining the actual resistance state in the tail region (useable resistance change, $\Delta R_{Use}$) of the distribution. For robust, less noise-sensitive, and high-speed read operation with normal process variation, large $\Delta R_{Use}$, and  significantly more than $12\sigma$ separation are essential \cite{toggle_mram2}. Furthermore, the width of resistance distribution varies from cell to cell because of manufacturing process variations. Besides, the quality, size, and level of in-homogeneity of the MTJ tunnel barrier have a significant impact on larger relative bit-to-bit resistance variation \cite{toggle_mram2, mtj_cell_res}. Therefore, a thicker tunnel barrier ($\sim1nm$) is essential to maintain the resistance level of the MTJ in the kilo $\Omega$ range for minimizing the series resistance effect from the isolation transistor \cite{toggle_mram2}, where $\Omega$ is the SI unit of resistance.

Storing data in a magnetic state has several benefits over charge-based storage such as non-destructive read operation, unlimited read/write endurance, no leakage during magnetic polarization, no wear-out due to no movement of electrons/atoms during the switching process of magnetic polarization \cite{toggle_mram3}, etc. Moreover, Savtchenko switching based MRAM arrays possesses several important performance characteristics such as lower write error rate and fast read/write cycle ($35 ns$). They are also less sensitive to external fields, and therefore they are less sensitive to manufacturing process variations \cite{toggle_engel}.  

For all commercial memory chips, manufacturers define a set of timing parameters for reliable read/write operation of the chips against a wide range of operating conditions. The write operation of the MRAM chip can be governed by three different control parameters: write enable ($\overline{W}$), chip enable ($\overline{E}$), and upper/lower byte enable ($\overline{UB}$/$\overline{LB}$) signals. A simplified version of the write enable ($\overline{W}$) controlled write operation of the MRAM chip is shown in Fig. \ref{fig:Everspin}.


Here,

$t_{WC}$ = \textit{write cycle time}, i.e., the time period to complete full write operation in a particular address.

$t_{W}$ = \textit{write~pulse~width}, i.e., the time period for which the $\overline{W}$ pin is kept activated. 

$t_{WR}$ = \textit{write~recovery~time}, i.e., the time to complete the write operation after the $\overline{W}$ pin is deactivated. 

$t_{DV}$ = \textit{valid~data~to~end~of write}, i.e., the time for which the valid data need to be available in the data I/O before the $\overline{W}$ pin is deactivated.  

If the output enable ($\overline{G}$) becomes active at the same time, or after $\overline{W}$ is activated, the output will remain in the high impedance state. After all three write control parameters ($\overline{E}$, $\overline{W}$, or $\overline{UB}$/$\overline{LB}$) become disabled, the $\overline{G}$ signal must remain in the steady-state high for at least $2 ns$. Reducing any of these timing parameters can improve the speed and reduce power consumption but may lead to faulty operation. The write timing parameter $t_{W}$ is manipulated in this work to introduce errors during $\overline{W}$ controlled write operation.

\section{Generating Random Numbers using COTS MRAM} \label{sec:method}

In our proposed methodology, we exploit the random Savtchenko switching at the reduced (manufacturer-recommended) timing parameter for generating true random numbers. When the \textit{write pulse width}, $t_{W}$, of toggle MRAM (see Fig.~\ref{fig:Everspin}) is reduced, it does not get sufficient time and write current to toggle into the desired stable state. Due to the process variation and the non-uniform distribution of current pulse within the chip, random variations are created in the MTJ storage element. Therefore, all of the memory cells are not capable of performing an appropriate write operation. That is the reason that a manufacturer specifies a set of timing parameters for reliable read/write operations. Violation in any of these manufacturer-recommended timing parameters may cause erroneous/faulty outputs during the read/write operation.
If the $t_{W}$ is not sufficient, there is a chance that FML is not aligned perfectly with the RML (either the same or in the opposite direction) and might settle on an intermediate position. This arrangement may lead the cell resistance to be halfway between $R_{Low}$ and $R_{High}$ \cite{TMR}. Therefore, at reduced $t_{W}$, if the resultant cell resistance falls around the $\Delta R_{Use}$ region of the resistance distribution (see Fig. \ref{fig:cell_res}) curve, the cell will show indeterministic characteristics and generate random bits.

In our proposed scheme, several steps are involved in generating true random numbers. At the reduced $t_{W}$, MRAM chips create errors, and the total number of errors differ at different reduced $t_{W}$ values.  At first, we select the most suitable reduced $t_{W}$ value. This selected $t_{W}$ aims to maximize the number of cells that can be used for TRNGs. Second, we propose a cell selection algorithm to characterize all of the temporally unbiased MRAM cells from a set of measurements to identify the most appropriate memory cells for generating robust random numbers. The mentioned two steps must be performed only once to choose the appropriate number of random MRAM cells. Finally, we collect data from multiple measurements from the selected MRAM cells and use a low-overhead post-processing technique to generate high-quality random numbers.

\subsection{Appropriate Reduced Time Selection}\label{subsec:red_time}

The experimental results show that some of the memory cells provide erroneous outputs if the data is written at the reduced timing parameters. The number of these error-prone cells varies within the write pulse activation time range t = [0,$t_{W}$]. We change the $t_{W}$ and count the total number of erroneous bit cells. Our main objective is to find a suitable $t_{W}$ for which the maximum number of erroneous bits is achieved. The number of erroneous cells is calculated from all achievable reduced write timing parameters in the next step. Finally, we propose an algorithm to characterize the memory cells among those erroneous cells to generate random numbers for any system using the timing parameter for which the maximum number of random cells is obtained.  Details about the cell characterization technique are described in Sect. \ref{subsec:cell_select}.

\subsection{MRAM Cells Characterization}\label{subsec:cell_char}

Our experimental result manifests that all of the memory cells are not suitable to generate robust random numbers. To locate these random cells, at first, we characterize MRAM memory cells by writing different intuitive (solid) and non-intuitive (random, checkerboard, and striped) input data patterns to the entire memory cells at the reduced write enable time, $t_{W}$, and read back the full memory contents with appropriate timing parameters a total of $N$ times. Larger $N$ provides better characterization results but increases the computation time for characterization. 

Theoretically, reduced write operation reduces the current flowing through the MTJ storage elements \cite{redu_write}. Hence, the magnetic orientation switching (\textit{parallel (P)} $\rightarrow$ \textit{anti-parallel (AP)} or vice versa) time increases significantly \cite{redu_write}. Switching from \textit{P} to \textit{AP} is more vulnerable to reduced write operation due to enhanced switching delay, leading to the write failure. Our experimental results also manifest that the write operation at the reduced $t_{W}$ produces erroneous data. Moreover, these error patterns depend on the input data pattern to be written and vary with different memory chips. Based on the error patterns from the sample measurements, the MRAM cells can be classified into the following two categories-

\textbf{Persistent  Cells:} These cells produce stable output from measurement to measurement. These stable cells are excellent candidates to generate memory-based Physically Unclonable Function \cite{mram_puf_trng, trng_khan} but not competent for true random number generation because of manifesting consistent behavior at the reduced $t_{W}$.

\textbf{Noise-prone Cells:} These cells provide inconsistent output for different measurements. However, we also observe that most noise-prone cells are biased toward `0' or `1'. Therefore, to avoid producing deterministic random numbers, we propose a cell selection technique to exclude those biased cells from the noise-prone cells (described in Sect. \ref{subsec:cell_select}).

\subsection{Appropriate Cell Location Selection}\label{subsec:cell_select}

To generate a robust TRNG, unbiased cells need to be filtered because all cells do not provide the same amount of entropy. At first, we discover all erroneous cells at the reduced $t_{W}$ from ($N$) measurements. At the reduced $t_{W}$, some cells will not create any errors; we define them as the correct state ($\mathcal{S}_{C}$). On the other hand, the other cells will create erroneous outputs; we define them as the error state ($\mathcal{S}_{E}$). 
Next, we record the change of state ($\mathcal{S}_{C} \rightarrow \mathcal{S}_{E}$ or $\mathcal{S}_{E} \rightarrow \mathcal{S}_{C}$) or flip of all cells comparing to the two consecutive measurements from each of $N$ measurements. This forms a $(1 \times M)$ array containing the total number of flips in each cell location from $N$ measurements, where $M$ is the total number of memory cells. Second, to select the random cells, we need to determine the appropriate threshold, $Th$. Theoretically, the expected value of $Th$ is $p \times (N-1)$, where $p$ ($ = \frac{1}{2}$) is the probability of state change (flip). However, in reality, a fixed $Th$ might not provide sufficient random cells. Hence a specified bound ($Th_L \le Th \le Th_U$) needs to be defined, where $Th_L$ and $Th_U$ are the lower- and upper-bound of the threshold range. Cells within this boundary are considered as TRNG candidates. The silicon results show that the obtained random cells above $Th_U$ are significantly negligible. Therefore, we only choose those cells for which the $(1 \times M)$ array contents are above $Th_L$. The locations of these unbiased noisy cells are stored in data-set $\mathcal{F}_{C}$. The step-by-step procedure is shown in Algorithm~\ref{alg:alg1}. A true random number must be highly temporal variant, which is the basis of our proposed algorithm. Therefore, the selected random cells with the proposed algorithm are capable of generating high-entropy random numbers.

\begin{algorithm}
 \caption{Pseudo-code for Random Cell Location Selection}
    \label{alg:alg1}
 \begin{algorithmic}[1]
 \renewcommand{\algorithmicrequire}{\textbf{procedure}}
 \renewcommand{\algorithmicensure}{\textbf{end}}
 \REQUIRE $rand\_cell\_loc(N, input\_data, Th_L)$
 \STATE /* $N$ = Number of total measurements */
 \STATE /* $input\_data$ = $(1 \times num\_cell)$ matrix containing input data stored to each memory cell at reduced $t_{W}$*/
 \STATE /* $Th_L$ = Lower threshold bound to choose true random cells */
 \STATE $num\_cell$ = Total number of memory cells 
 \STATE /* $flip\_count$ matrix stores total number of state change (flip) from consecutive N measurements */
 \STATE $flip\_count = zeros(1 \times num\_cell)$ 
 \FOR {$i = 1$ to $N - 1$}
 \STATE $x = bitwise\_xor(input\_data(i), input\_data(i+1))$
 \STATE $flip\_count = bitwise\_add (x, flip\_count)$
 \ENDFOR
 \STATE /* $rand\_loc$ matrix stores random cell locations */
 \STATE $rand\_loc = zeros(1 \times num\_cell)$
 \STATE $num\_randcell = 0$ /* Total number of random cells */
 \FOR {$i = 1$ to $num\_cell$}
 \IF {$flip\_count(i) \ge Th_L$}
 \STATE $rand\_loc(i) = 1$
 \STATE $num\_randcell++$
 \ENDIF
 \ENDFOR
 \RETURN $rand\_loc$
 \ENSURE procedure
 \end{algorithmic}
 \end{algorithm}
 \squeezeupf

\subsection{Low-overhead Post-processing}\label{subsec:post_proc}

However, the raw random sequence that apparently seems unbiased might provide biased results under extreme operating conditions. Therefore, several post-processing techniques such as Von Neumann corrector, XORing multiple bits, cryptographic hash function, etc. are used to generate a fully non-deterministic random sequence \cite{post_proc}. Secure Hash Algorithm (SHA)-256 is area efficient, fast, and fewer input bits are required to generate the same entropy level \cite{intel, tpm}. Therefore, we chose SHA-256 as a post-processing technique and applied over the bit sequence obtained from $\mathcal{F}_{C}$. To generate a random number of the required length, at first, we accumulate the obtained random cells at the selected reduced $t_{W}$. The value of the measurements is a function of the required length and the used post-processing technique. Next, the bit sequence is split into appropriate chunks to feed into the SHA-256 hash algorithm. Finally, the multiple output chunks gathered from the SHA-256 hash function are concatenated to produce truly unbiased random numbers and are denoted as $\mathcal{H}_{C}$.

\section{Experimental Results and Analysis} \label{sec:result}

The primary analysis is performed over ten (2 chips from each \textit{MR0A16ACYS35, MR0A16AYS35, MR1A16AYS35 MR2A16ACYS35, MR2A16AYS35} models) 16-bit parallel interfaced differently sized ($1 Mb - 4 Mb$) memory chips manufactured by Everspin technologies. Among them, five chips are selected randomly to perform extensive analysis for TRNG.  We have used our own custom memory controller implemented on \textit{Xilinx Artix 7  (XC7A35T-1C)} FPGA to manipulate different timing latency of a couple of emerging memories \cite{FPGA}. As discussed in Sect. \ref{subsec:cell_char}, the generated error is pattern dependent at reduced operation. Hence to determine the suitable pattern that is capable of generating high-entropy true random numbers, we collected a total of 5-set measurement data with seventeen different intuitive (solid) and non-intuitive (random, checkerboard, and striped) 16-bit input data patterns: ($0xFFFF$, $0xAAAA$, $0x5555$, $0x0000$) from each ten memory chips. We observed that the \textit{solid} $0x0000$ pattern produces comparatively high erroneous bits than other patterns. Therefore, we can conclude that \textit{parallel (anti-parallel)} configuration is the logic state ‘1’ (state ‘0’). Next, to characterize the MRAM cells (discussed in Sect. \ref{subsec:cell_char}), we collected a total of 50-set measurement data with only \textit{solid} $0x0000$ input pattern from the selected five memory chips. We chose the smallest possible achievable (due to experimental limitation) value of $t_W$, $16.6\%$ of the recommended $t_W$, for this work. However, our selected value of $t_W$ can generate a sufficient number of incorrect outputs to generate high-quality random numbers.  

\subsection{Selection of \texorpdfstring{$t_W$}{Lg}}\label{subsec:res_A}

To compare the behavior of the faulty/erroneous outputs at different reduced $t_W$ values using \textit{solid} $0x0000$ data pattern, an analysis is performed to determine the cell types (i.e., noisy or persistent). We reduce the $t_W$ value from $15 ns$ (manufacturer's recommended) to  $10 ns$, $5 ns$, and $2.5 ns$, respectively. Due to the experimental set-up limitations, we are incapable of reducing the $t_W$ value any further. At $t_W$ = $5 ns$ and $10 ns$, the obtained total failed bits are almost negligible ($< 5\%$ and $< 1\%$, respectively) for all ten chips. However, at $t_W$ = $2.5 ns$, the total number of failed bit count falls within $25.59\% - 37.30\%$, which is sufficient for TRNG analysis. Hence, we choose $t_W$ = $2.5 ns$ (considering the number of failed bit count) to characterize erroneous cells.  

\subsection{Characterization of Temporally Unbiased Cells}\label{subsec:res_B}

The MRAM cell characterization is performed according to Sect. \ref{subsec:cell_select} with $N = 50$ measurements. Table \ref{Tab:summary} shows a summary of random MRAM addresses and cells after performing cell characterization. The results show that the total number of random cells obtained at reduced $t_W$ ($2.5 ns$) varies from chip to chip. The results also show that different memory modules may have different thresholds. We also observe that only a few addresses hold these random cells. 
In Table \ref{Tab:summary}, the first row shows the cell selection threshold, $Th_L$, used for different chips. 
As we need to perform characterization so different thresholds for different models will be acceptable- the reason for choosing different thresholds to ensure higher throughput. However, the same threshold is used for the same model, i.e., C1 \& C2 are from the same model. We can determine a unique threshold for simplicity for all models considering the highest threshold value (in our case, 23); however, at that point, we will get lower throughput as a lower threshold provides higher throughput (see Sect. \ref{subsec:res_F}).
The second row represents the percentage of addresses that contain random cell(s). Note that we only consider those addresses which have at least one cell that lies into $\mathcal{F}_{C}$. As the percentage of random addresses is very small ($\sim 1\%$) regardless of memory size, we will need to store only those few memory cells' information. Finally, the last row presents the average number of random bits per random addresses.

\begin{table}[ht!]
\caption{Cell statistics after applying cell characterization algorithm.}
\setcellgapes{2pt}
\captionsetup{justification=centering, margin= 0cm}
\makegapedcells
\centering
\setlength\tabcolsep{2pt} 
\resizebox{0.41\textwidth}{!}
{
    \begin{tabular}{|c|c|c|c|c|c|}
    \hline
    Sample   Chip\footnotemark    & C1 & C2 & C3 & C4 & C5 \\ \hline
    Threshold ($Th_L$)            & 16     & 16       & 15     & 23     & 21     \\ \hline
    \#(Rand Addr) (\%)            & 1.16   & 1.5    & 1.23   & 0.63   & 0.94   \\ \hline
    \#(Rand Bits)$/$\#(Rand Addr) & 9.71   & 10.71  & 13.19  & 11.39  & 12.76  \\ \hline
    \end{tabular}
}
\label{Tab:summary}
\vspace{-6mm} 
\end{table}
\footnotetext{C1\&C2: MR0A16ACYS35, C3: MR0A16AYS35, C4: MR2A16ACYS35, C5: MR2A16AYS35}
\begin{figure}[htbp]
    \centering
    \captionsetup{justification=centering, margin= 0.5cm}
    \includegraphics[trim=0.5cm 0cm 1.5cm 1cm, clip, width = 0.35\textwidth]{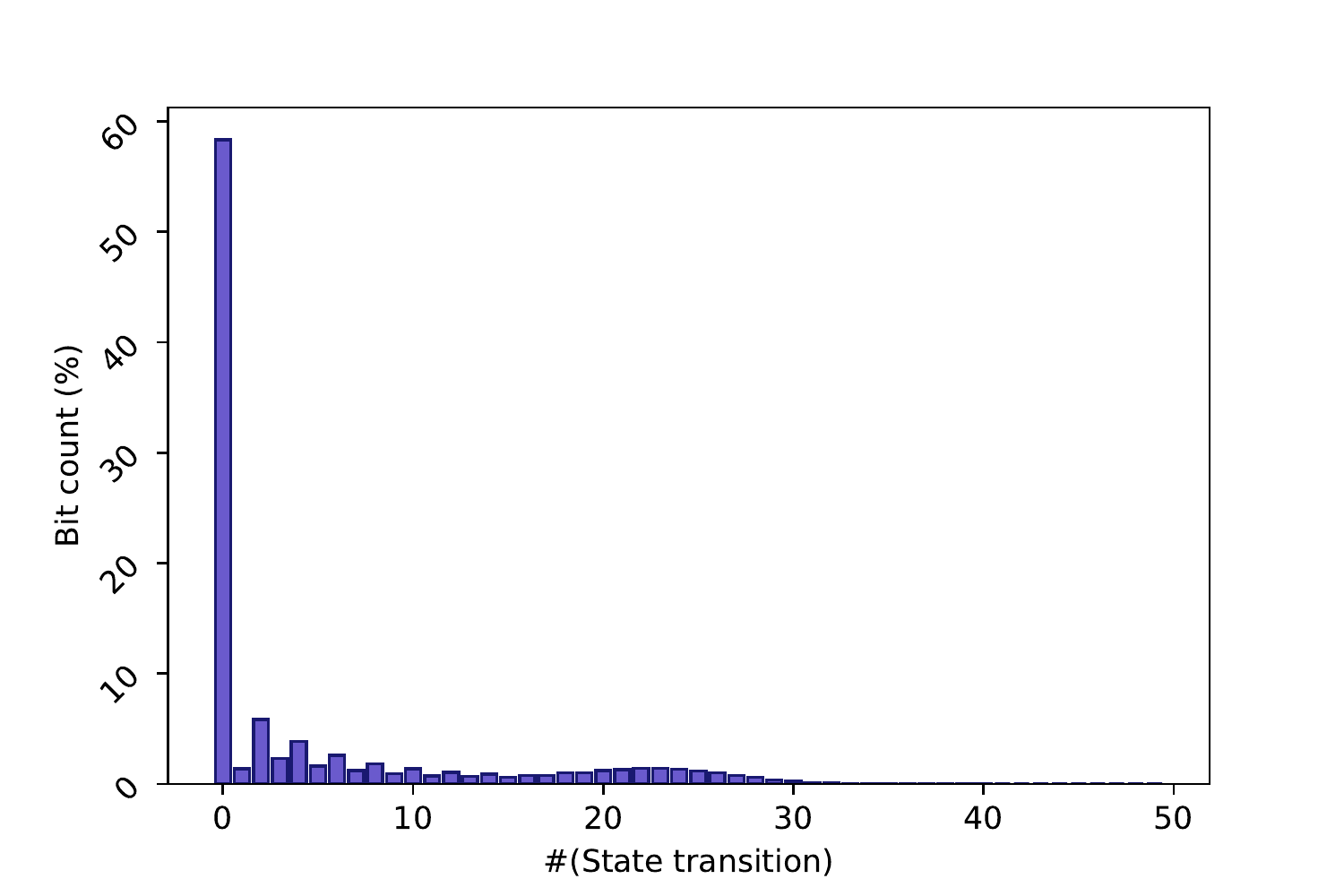}
    \caption{The characteristics of memory cells of \textit{C4}: most of the cells are purely invariant (stuck at `0'$/$`1').}
    \label{fig:MR4_ch2}
\end{figure}
%

Fig. \ref{fig:MR4_ch2} illustrates the number of flips ($\mathcal{S}_{C} \rightarrow \mathcal{S}_{E}$ or $\mathcal{S}_{E} \rightarrow \mathcal{S}_{C}$) of memory cells with $N = 50$ temporal measurements for a randomly chosen memory chip. \textit{Solid} $0x0000$ is used as the write data pattern. Note that to erase the trace of previously written data, we reset the entire memory with the \textit{solid} $0xFFFF$ data pattern before every measurement. The results show that a larger number of cells ($\sim 40 - 60\%$) are purely invariant (stuck at `0'$/$`1'logic state), which is not desirable for high-entropy random number generator. To filter out those temporally persistent cells using the proposed algorithm (described in Sect. \ref{subsec:cell_select}), we only chose those cells (set $\mathcal{F}_{C}$) that are above the lower threshold range $Th_L$ = $[15, 23]$ (see Table \ref{Tab:summary}) for different chips used in the experiment. Although the number of eligible cells decreases after performing our proposed cell selection algorithm, the filtered cells are enough to generate high-quality random numbers.

\subsection{Evaluation}\label{subsec:res_C}

\begin{table*}[ht!]
\caption{The worst-case NIST test results at the reduced $t_W$.}
\setcellgapes{1pt}
\captionsetup{justification=centering, margin= 0cm}
\makegapedcells
\centering
\setlength\tabcolsep{5pt} 
\resizebox{0.95\textwidth}{!}
{
    \begin{tabular}{|c|c|c|c|c|c|c|c|c|c|c|}
    \hline
    Sample Chip &
      \multicolumn{2}{c|}{C1} &
      \multicolumn{2}{c|}{C2} &
      \multicolumn{2}{c|}{C3} &
      \multicolumn{2}{c|}{C4} &
      \multicolumn{2}{c|}{C5} \\ \hline
    Result Type &
      $\mathcal{P}-val.$ &
      $Prop.$ &
      $\mathcal{P}-val.$ &
      $Prop.$ &
      $\mathcal{P}-val.$ &
      $Prop.$ &
      $\mathcal{P}-val.$ &
      $Prop.$ &
      $\mathcal{P}-val.$ &
      $Prop.$ \\ \hline
    Frequency               & 0.122325 & 14/14 & 0.637119 & 20/20 & 0.141256 & 21/21 & 0.534146 & 11/12 & 0.991468 & 20/20 \\ \hline
    BlockFrequency          & 0.350485 & 14/14 & 0.350485 & 20/20 & 0.980883 & 21/21 & 0.534146 & 12/12 & 0.834308 & 20/20 \\ \hline
    CumulativeSums          & 0.066882 & 14/14 & 0.275709 & 20/20 & 0.105618 & 21/21 & 0.350485 & 11/12 & 0.534146 & 20/20 \\ \hline
    Runs                    & 0.213309 & 14/14 & 0.637119 & 20/20 & 0.311542 & 21/21 & 0.534146 & 12/12 & 0.437274 & 20/20 \\ \hline
    LongestRun              & 0.066882 & 14/14 & 0.534146 & 19/20 & 0.392456 & 21/21 & 0.739918 & 12/12 & 0.122325 & 20/20 \\ \hline
    Rank                    & 0.534146 & 14/14 & 0.275709 & 19/20 & 0.875539 & 21/21 & 0.739918 & 12/12 & 0.534146 & 20/20 \\ \hline
    FFT                     & 0.534146 & 14/14 & 0.275709 & 20/20 & 0.105618 & 21/21 & 0.350485 & 12/12 & 0.534146 & 19/20 \\ \hline
    NonOverlappingTemplate  & 0.017912 & 12/14 & 0.122325 & 18/20 & 0.311542 & 19/21 & 0.213309 & 11/12 & 0.350485 & 18/20 \\ \hline
    OverlappingTemplate     & 0.213309 & 14/14 & 0.275709 & 20/20 & 0.021262 & 21/21 & 0.534146 & 12/12 & 0.275709 & 20/20 \\ \hline
    Universal               & 0.017912 & 14/14 & 0.739918 & 20/20 & 0.311542 & 21/21 & 0.213309 & 12/12 & 0.637119 & 18/20 \\ \hline
    ApproximateEntropy      & 0.035174 & 14/14 & 0.637119 & 20/20 & 0.585209 & 21/21 & 0.213309 & 12/12 & 0.017912 & 20/20 \\ \hline
    RandomExcursions        & 0.991468 & 10/12 & 0.213309 & 11/12 & 0.534146 & 11/12 & ----     & 7/7   & 0.006196 & 13/13 \\ \hline
    RandomExcursionsVariant & 0.017912 & 11/12 & 0.017912 & 11/12 & 0.122325 & 11/12 & ----     & 6/7   & 0.048716 & 12/13 \\ \hline
    Serial                  & 0.350485 & 13/14 & 0.739918 & 19/20 & 0.105618 & 21/21 & 0.350485 & 12/12 & 0.437274 & 19/20 \\ \hline
    LinearComplexity        & 0.350485 & 14/14 & 0.534146 & 20/20 & 0.689019 & 21/21 & 0.739918 & 12/12 & 0.534146 & 20/20 \\ \hline
    \multicolumn{4}{l}{$^{\mathrm{*}}$NB. —- test not performed due to insufficient data \cite{NIST}.}
    \end{tabular}
}
\label{Tab:NIST1}
\vspace{-6mm}
\end{table*}

Data collected from different FPGAs verify that the memory controllers do not influence the randomness of the generated random number from MRAM chips. Furthermore, to evaluate the quality, randomness, and effectiveness of the obtained binary sequence, set $\mathcal{H}_{C}$ (after performing low-overhead post-processing technique, as discussed in Sect. \ref{subsec:post_proc}, over the test data sequence), the most frequently used and well-accepted NIST statistical test suite (STS) \cite{NIST} is used. Tables \ref{Tab:NIST1} and \ref{Tab:NIST2} show the worst-case (i.e., from multiple similar test categories, the worst one is exhibited) NIST test results considering different memory models and extreme operating conditions. In the tables, the higher p-value ($\mathcal{P}-val.$) (calculated from the chi-squared ($\chi^{2}$) test) indicates a purely random sequence and vice versa. Besides, $Prop.$ is the proportion of the binary sequence that passes the corresponding test. However, for passing the randomness test, the minimum value of $\mathcal{P}-val.$ should be $0.0001$, and $Prop.$ needs to be greater than a specified value, which depends on the number of sample sizes (i.e., minimum of 18 tests need to be passed for 20 binary test sequences). The proposed MRAM-based binary sequences pass all (15) of the NIST tests; thus, it can be considered purely random.

\begin{table*}[ht!]
\caption{The worst-case NIST test results verify the robustness of our proposed TRNG.}
\setcellgapes{2pt}
\captionsetup{justification=centering, margin= 0cm}
\makegapedcells
\centering
\setlength\tabcolsep{2pt} 
\resizebox{0.95\textwidth}{!}
{
    \begin{tabular}{|c|c|c|c|c|c|c|c|c|c|c|c|c|}
    \hline
    Sample Chip & \multicolumn{6}{c|}{C3}   & \multicolumn{6}{c|}{C4}                \\ \hline
    Operating Condition &
      \multicolumn{2}{c|}{$T_{High} (65^{\circ}C)$} &
      \multicolumn{2}{c|}{$T_{Low} (20^{\circ}C)$} &
      \multicolumn{2}{c|}{M-Field $(8mT)$} &
      \multicolumn{2}{c|}{$T_{High} (65^{\circ}C)$} &
      \multicolumn{2}{c|}{$T_{Low} (20^{\circ}C)$} &
      \multicolumn{2}{c|}{M-Field $(8mT)$} \\ \hline
    Result Type             & $\mathcal{P}-val.$  & $Prop.$ & $\mathcal{P}-val.$  & $Prop.$ & $\mathcal{P}-val.$  & $Prop.$ & $\mathcal{P}-val.$  & $Prop.$ & $\mathcal{P}-val.$  & $Prop.$ & $\mathcal{P}-val.$  & $Prop.$ \\ \hline
    Frequency               & 0.788728 & 21/21 & 0.585209 & 21/21 & 0.242986 & 21/21 & 0.739918 & 12/12 & 0.534146 & 12/12 & 0.122325 & 11/12 \\ \hline
    BlockFrequency          & 0.311542 & 20/21 & 0.585209 & 21/21 & 0.392456 & 21/21 & 0.534146 & 12/12 & 0.739918 & 11/12 & 0.066882 & 12/12 \\ \hline
    CumulativeSums          & 0.078086 & 21/21 & 0.105618 & 21/21 & 0.311542 & 21/21 & 0.213309 & 12/12 & 0.002043 & 12/12 & 0.017912 & 10/12 \\ \hline
    Runs                    & 0.689019 & 21/21 & 0.875539 & 21/21 & 0.392456 & 21/21 & 0.122325 & 12/12 & 0.739918 & 12/12 & 0.213309 & 12/12 \\ \hline
    LongestRun              & 0.311542 & 21/21 & 0.242986 & 21/21 & 0.242986 & 20/21 & 0.213309 & 12/12 & 0.008879 & 12/12 & 0.066882 & 12/12 \\ \hline
    Rank                    & 0.311542 & 21/21 & 0.186566 & 21/21 & 0.057146 & 21/21 & 0.122325 & 11/12 & 0.534146 & 12/12 & 0.534146 & 12/12 \\ \hline
    FFT                     & 0.585209 & 21/21 & 0.392456 & 21/21 & 0.186566 & 21/21 & 0.911413 & 12/12 & 0.000439 & 11/12 & 0.122325 & 12/12 \\ \hline
    NonOverlappingTemplate  & 0.311542 & 19/21 & 0.186566 & 19/21 & 0.242986 & 19/21 & 0.008879 & 11/12 & 0.017912 & 11/12 & 0.122325 & 11/12 \\ \hline
    OverlappingTemplate     & 0.242986 & 21/21 & 0.392456 & 20/21 & 0.585209 & 21/21 & 0.350485 & 12/12 & 0.213309 & 12/12 & 0.008879 & 11/12 \\ \hline
    Universal               & 0.689019 & 20/21 & 0.875539 & 20/21 & 0.242986 & 21/21 & 0.350485 & 11/12 & 0.739918 & 12/12 & 0.534146 & 12/12 \\ \hline
    ApproximateEntropy      & 0.141256 & 20/21 & 0.141256 & 20/21 & 0.029796 & 21/21 & 0.534146 & 12/12 & 0.213309 & 11/12 & 0.911413 & 12/12 \\ \hline
    RandomExcursions        & 0.004301 & 14/14 & 0.066882 & 12/12 & 0.066882 & 11/12 & ----     & 9/9   & ----     & 6/7   & ----     & 5/5   \\ \hline
    RandomExcursionsVariant & 0.213309 & 13/14 & 0.066882 & 11/12 & 0.213309 & 11/12 & ----     & 9/9   & ----     & 6/7   & ----     & 4/5   \\ \hline
    Serial                  & 0.392456 & 21/21 & 0.392456 & 20/21 & 0.186566 & 21/21 & 0.350485 & 12/12 & 0.213309 & 11/12 & 0.350485 & 12/12 \\ \hline
    LinearComplexity        & 0.311542 & 20/21 & 0.057146 & 21/21 & 0.186566 & 21/21 & 0.534146 & 11/12 & 0.035174 & 12/12 & 0.213309 & 12/12 \\ \hline
    \multicolumn{4}{l}{$^{\mathrm{*}}$NB. —- test not performed due to insufficient data \cite{NIST}.}
\end{tabular}
}
\label{Tab:NIST2}
\squeezeupf 
\end{table*}
%

\subsection{Robustness Analysis}\label{subsec:res_D}
\noindent
\textbf{Chip Variations:} 
The silicon results from four different memory models of two different sizes show that our proposed TRNG is robust. However, the statistics of random cells are different for different memory models, shown in Table \ref{Tab:summary}. These sources of variations come from architectural as well as both inter- and intra-chip dissimilarities. As the random process variation is the key source of any memory chips' randomness, the proposed scheme can generate random numbers.

\noindent
\textbf{Environmental Variations:} Robustness against a wide range of operating conditions is one of the requirements of high-quality TRNGs. To verify the robustness of our proposed random number under temperature variation and external magnetic field (M-Field), we collected four sets of test data sequence at different operating conditions: i) room temperature ($26^{\circ}C$), ii) high temperature ($65^{\circ}C$), and iii) low temperature ($20^{\circ}C$) without external M-Field. The fourth set is collected at room temperature with an $\sim 8mT$ external M-Field. 
The temperature range is chosen within the manufacturer's recommended value, which is $[ 0^{\circ}C - 70^{\circ}C]$ for all memory models. 
The total number of random cells is observed comparatively less for low temperatures than the other operating conditions. The write latency of MRAM increases significantly at the lower temperature, which results in the reduction of the number of random cells at the reduced write operation \cite{mtj_cell_res}. Besides, we applied a constant rare earth magnetic source (generated from the permanent magnet) in six different orientations of 3D coordinates to observe the effect of external M-Field. However, we did not notice any significant change in the total number of random cells with ($8mT$) external M-Field. Therefore, we can conclude that our proposed random numbers are robust against extreme operating conditions. To further evaluate the robustness of the cell selection threshold, we deliberately select those two models (\textit{C3} $\&$ \textit{C4}) with the lowest and highest $Th_L$ values (see Tables \ref{Tab:summary} and \ref{Tab:NIST2}). Table~\ref{Tab:NIST2} shows the worst-case NIST STS results in extreme operating conditions, signifying that our proposed TRNG can produce high-quality random numbers.

\subsection{Throughput Analysis}\label{subsec:res_F}

Cell characterization and registration are performed once during a full life cycle. Therefore, the registration phase is not considered during throughput calculation. The TRNG throughput of our proposed technique is the function of the average time required to perform read/write operation from one memory cell at the reduced $t_W$ and the average time required to execute the SHA-256 hash function. The throughput of our proposed TRNG is calculated as follows:

\begin{equation}\label{eqn:throughput}
    \mathcal{T}_{avg, worst} = \frac{\mathcal{D}_{len}}{\boldsymbol{t}_{RW,avg} + \boldsymbol{t}_{hash,avg}}
\end{equation}
\squeezeupt
where,
\begin{align} 
    \boldsymbol{t}_{RW,avg}= \frac{\boldsymbol{t}_{RW} \times \mathcal{B}_{len}}{(\frac{\#Rand Bits}{\#Rand Addr})}
\end{align}

Here, $\mathcal{B}_{len} (= 512$\textit{-bit)} and $\mathcal{D}_{len} (= 256$\textit{-bit)} are the length of the input and hashed output (message digest) block size of the hash function, respectively, $\boldsymbol{t}_{hash,avg}$ is the average time required to hash the input bit sequence of length $\mathcal{B}_{len}$,  $\boldsymbol{t}_{RW,avg}$ is the average time required to generate raw random bits of size $\mathcal{B}_{len}$, $\boldsymbol{t}_{RW}$ is the average time required to perform a complete read/write operation from one memory address, and `\textit{\#(Rand Bits)$/$\#(Rand Addr)}' (see Table \ref{Tab:summary}) is the average number of random bits per random addresses. The cryptographic hash function, SHA-256, is used in this work due to the low-overhead post-processing. Nowadays, almost all modern processors have dedicated instruction set architecture to provide hardware support for performing the secure hash algorithm \cite{intel}. We found $\boldsymbol{t}_{hash,avg} = 802.6 ns$ using \textit{Intel i7-8700} processor. Note that we use a high-level language (\textit{Python API}) to hash the complete $512$-bit block message; hence, the obtained average time ($802.6 ns$) includes the function overhead. Ideally, MRAM has a comparatively fast ($35 ns$) read/write cycle considering the nominal operation \cite{everspin}. However, using our evaluation board, the obtained $\boldsymbol{t}_{RW}$ considering reduced write operation is $239.76 ns$ (much higher than $70 ns$), which signifies the inclusion of the communication overhead between the FPGA interfaced with a computer to acquire the data from memory for analysis. Furthermore, Table \ref{Tab:summary} shows the different `\textit{\#(Rand Bits)$/$\#(Rand Addr)}' values for different memory chips. According to Eq. \ref{eqn:throughput}, our system-level worst-case throughput values are around $18.17,~19.95,~24.12,~21.10$, and $23.47 Mbit/s$ for $C1 - C5$ chips, respectively, considering all of the communication and function overhead. The obtained throughput values are significantly higher compared with the performance of many popular (non) volatile memory-based TRNGs \cite{trng_dram_kim, FRAM, trng_flash}. An efficient implementation of a memory controller can further improve the overall performance of our proposed TRNG.

\section{Conclusion} \label{sec:end}

This paper demonstrated an efficient technique to generate high-throughput and high-quality true random numbers from non-volatile COTS MRAM chips by utilizing its internal write latency variation. The NIST SP-800-22 suite results validate that our proposed technique is purely random and robust at extreme operating conditions. The throughput is also considerably higher than most of the available TRNG techniques implemented using existing or emerging volatile or NVM chips.


\vspace{12pt}


\begin{thebibliography}{00}

\bibitem{trng_Mathew}
S. K. Mathew et al., “2.4 Gbps, 7 mW all-digital PVT-variation tolerant true random number generator for 45 nm CMOS high-performance microprocessors,” IEEE J. Solid-State Circuits, vol. 47, no. 11, pp. 2807–2821, 2012.


\bibitem{trng_dram_kim}
J. S. Kim et al., “D-RaNGe: Using Commodity DRAM Devices to Generate True Random Numbers with Low Latency and High Throughput,” IEEE International Symposium on HPCA, pp. 582-595, 2019.

\bibitem{trng_kim}
E. Kim, M. Lee, and J.-J. Kim, “8.2 8Mb/s 28Mb/mJ robust true-random-number generator in 65nm CMOS based on differential ring oscillator with feedback resistors,” IEEE International Solid-State Circ. Conf., 2017. 
\bibitem{TRNG_rahman_sram}
M. T. Rahman et al., “Enhancing noise sensitivity of embedded SRAMs for robust true random number generation in SoCs.” IEEE Asian Hardware-Oriented Security and Trust (AsianHOST), 2016.

\bibitem{trng_yang}
K. Yang et al., “16.3 a 23mb/s 23pj/b fully synthesized true-random-number generator in 28 nm and 65 nm CMOS,” in IEEE Int. Solid-State Circuits Conf. Dig. Tech. Papers, vol. 57, pp. 280–281, 2014.

\bibitem{trng_mStability}
C. Tokunaga,  D. Blaauw, and T. Mudge, “True Random Number Generator with a MetastabilityBased Quality Control,” ISSCC Dig. Tech. Papers, pp. 404-405, 2007. 

\bibitem{rtn_Brederlow}
R. Brederlow et al., “A low-power true random number generator using random telegraph noise of single oxide-traps,” in IEEE Int. Solid-State Circuits Conf. (ISSCC) Dig. Tech. Papers, pp. 1666–1675, 2006.

\bibitem{trng_reram}
C.-Y. Huang et al., “A contact-resistive random-access-memory-based true random number generator,” IEEE Electron Device Lett., vol. 33, no. 8, pp. 1108–1110, 2012.

\bibitem{TRNG_rahman} 
M. T. Rahman et al., “TI-TRNG: Technology independent true random number generator.” 51st ACM/EDAC/IEEE Design Automation Conference (DAC). IEEE, 2014.

\bibitem{trng_quantum}
C. Gabriel et al., ``A generator for unique quantum random numbers based on vacuum states,” Nature Pho., vol. 4, no. 10, pp. 711–715, 2010. 

\bibitem{Spin_ghosh}
S. Ghosh, ``Spintronics and Security: Prospects, Vulnerabilities, Attack Models, and Preventions,” Proceedings of the IEEE, vol. 104, no. 10, pp. 1864–1893, 2016.

\bibitem{mram_trng_choi}
W. H. Choi et al., ``A magnetic tunnel junction based true random number generator with conditional perturb and real-time output probability tracking,” in IEDM Tech. Dig., pp. 12.5.1–12.5.4, 2014.

\bibitem{Spin_chun}
S. Chun et al., ``High-density physical random number generator using spin signals in multidomain ferromagnetic layer,” Adv. Condens. Matter Phys., vol. 2015, Art. no. 251819, 2015.

\bibitem{reram_trng_wei}
Z. Wei et al., ``True random number generator using current difference based on a fractional stochastic model in 40-nm embedded ReRAM,” in IEDM Tech. Dig., pp. 4.8.1–4.8.4, Dec. 2016.

\bibitem{reram_trng_Balatti}
S. Balatti et al.,  ``True random number generation by variability of resistive switching in oxide-based devices,” IEEE J. Emerg. Sel. Topics Circuits Syst., vol. 5, no. 2, pp. 214–221, 2015.

\bibitem{trng_sara}
B. M. S. Bahar Talukder et al., “Exploiting DRAM latency variations for generating true random numbers,” in Proc. IEEE Int. Conf. Consum. Electron. (ICCE), pp. 1–6, Jan. 2019.


\bibitem{proc_var}
J. Lorenz et al., ``Simultaneous simulation of systematic and stochastic process variations,” in Proc. Int. Conf. SISPAD, pp. 289–292, 2014.

\bibitem{DRAM_limit}
J. A. Mandelman et al., ``Challenges and future directions for the scaling of dynamic random-access memory (DRAM),” IBM Journal of Research and Development, vol. 46, no. 2.3, pp. 187–212, 2002.

\bibitem{MRAM_propty}
T. Kishi et al., ``Lower-current and fast switching of a perpendicular TMR for high speed and high density spin-transfer-torque MRAM,” in IEDM Tech. Dig., pp. 1–4, 2008.




\bibitem{trng_flash}
Y. Wang et al., ``Flash Memory for Ubiquitous Hardware Security Functions: True Random Number Generation and Device Fingerprints," IEEE Symposium on Security and Privacy, pp. 33-47, 2012.


\bibitem{mram_trng_yang}
K. Yang et al., “A 28NM Integrated True Random Number Generator Harvesting Entropy from MRAM,” IEEE Symp. on VLSI Cir., 2018. 

\bibitem{trng_reram2}
H. Jiang et al., “A novel true random number generator based on a stochastic diffusive memristor,” Nature Commun., vol. 8, pp. 882, 2017.

\bibitem{FRAM}
M. I. Rashid et al., “True Random Number Generation Using Latency Variations of FRAM." IEEE Transactions on Very Large Scale Integration (VLSI) Systems, vol. 29, no. 1, pp 14-23, 2020.

\bibitem{trng_khan}
M. N. I. Khan et al., ``A Morphable Physically Unclonable Function and True Random Number Generator using a Commercial Magnetic Memory,” 21st IEEE International Symposium on Quality Electronic Design, 2020.

\bibitem{mram_puf_trng}
E. I. Vatajelu, G. D. Natale, and P. Prinetto, ``Security primitives (PUF and TRNG) with STT-MRAM,” IEEE 34th VLSI Test Symposium, 2016. 

\bibitem{mram_trng_Fukushima}
A. Fukushima et al., ``Spin dice: A scalable truly random number generator based on spintronics," JAP Express, 2014

\bibitem{mram_trng_Carboni}
R. Carboni et al., “Random Number Generation by Differential Read of Stochastic Switching in Spin-Transfer Torque Memory,” IEEE Electron Device Letters, vol. 39, no. 7, pp. 951–954, 2018.

\bibitem{everspin}
Everspin Parallel Interface MRAM. [Online]. Available: \url{https://www.everspin.com/parallel-interface-mram}. [Accessed: 05-Nov-2020].

\bibitem{toggle_mram}
L. Savtchenko et al., “Method of writing to scalable magnetoresistance random access memory element,” US Patent 6,545,906 B1, 2003.

\bibitem{toggle_mram2}
D. Apalkov et al., “Magnetoresistive Random Access Memory,” Proceedings of the IEEE, vol. 104, no. 10, pp. 1796–1830, 2016. 



\bibitem{mtj_cell_res}
J. M. Slaughter, “Materials for Magnetoresistive Random Access Memory,” Annu. Rev. Mater. Res., vol. 39, no. 1, pp. 277–296, 2009

\bibitem{toggle_mram3}
M. Durlam et al, “Toggle MRAM: A highly-reliable Non-Volatile Memory,” International Symp. on VLSI Tech., Systems and App., 2007. 

\bibitem{toggle_engel}
B. N. Engel, et al, “A 4-Mb toggle MRAM based on a novel bit and switching method,” IEEE Trans. Magn., vol. 41, pp. 132–136, 2005.

\bibitem{TMR}
“TMR Sensors”, Tech Notes, TDK Product Center,
{\url{https://product.tdk.com/info/en/products/sensor/angle/tmr_angle/technote/tpo/index.html}}.

\bibitem{redu_write}
R. Bishnoi et al., “Avoiding unnecessary write operations in STT-MRAM for low power implementation,” Int. Symp. on Quality Elec. Des., 2014.


\bibitem{post_proc}
S.-H. Kwok et al., “A comparison of post-processing techniques for biased random number generators, in IFIP Int. Workshop on Inf.n Sec. Theory and Practices. Springer, 2011, pp. 175-190.

\bibitem{intel}
J. Guilford, K. Yap, and V. Gopal, “Fast SHA-256 Implementations on Intel Architecture Processors, Intel white paper, May 2012. 

\bibitem{tpm}
D. Wooten, “Trusted platform module security,” U.S. Patent 9 230 109 B2, 2016.



\bibitem{FPGA}
Alchitry Au FPGA development board. [Online]. Available: {\url{https://alchitry.com/products/alchitry-au-fpga-development-board}}. 

\bibitem{NIST}
L. Bassham et al., “A Statistical Test Suite for Random and Pseudorandom Number Generators for Cryptographic Applications,” NIST, 2010.




\end{thebibliography}
\end{document}